\documentstyle{l-aa}
\begin{document}

\title{Long-term infrared variation of 3C279 and 4C29.45}

\author{J.H. Fan \inst{1,2} }

\institute{ CCAST (World Laboratory), P.O. Box 8730, Beijing
100080, China \and Center for
Astrophysics, Guangzhou Normal University, 
Guangzhou 510400, China, e-mail: 
jhfan@guangztc.edu.cn}

\date{Received <date>;accepted<date>}

\maketitle
\begin{abstract}
 In this paper, the historical infrared (JHK) data compiled from 22 
 publications are presented for 3C279 and 4C29.45. 
 Relations between color index and magnitude and between color-color 
 indices are discussed.  (J-K) is found closely
 correlated with (J-H) while (J-H) is almost not correlated with (H-K),
 which is perhaps due to two facts of (1) (J-K) shows wider distribution
 than (J-H) and (H-K) and (2) the spectrum deviates from the power law.
 For color-magnitude relation, there is a correlation of  color index
 increasing with the magnitude for 3C279 suggesting that the spectrum flattens
 when the source brightens. But an opposite tendency is also noted between
 (H-K) and J when J is fainter than 14 magnitude, which suggests that
 the near-IR spectrum of 3C 279 consists of, at least, two
components. No similar result is found for 4C29.45. The largest
 amplitude infrared variation is smaller than but comparable with the 
 largest amplitude variation for the two objects.

\end{abstract}

\section{Introduction}

 Blazars are AGNs characterized by compact radio core, high and variable 
 radio and optical polarization, superluminal radio components. The 
 continuum emissions are rapidly variable at all frequencies with
 amplitude of variability increasing with frequency (see Kollgaard 1994, 
 Urry \& Padovani 1995 and Scarp \& Falomo 1997). Blazars include BL 
 Lac objects, optically violently variable quasars (OVVs), highly
 polarized quasars(HPQs), flat spectrum radio quasars (FSRQs), and core 
 dominated  quasars (CDQ). All those objects are basically the same 
 thing (Fugmann  1989; Impey et al. 1991; Valtaoja et al. 1992; 
 Will et al. 1992; Scarpa \& Falomo 1997). 

 3C279 ( PKS 1253-055, 4C-05.55) and 4C29.45 (QSO 1156+295, Ton 599) 
 are members of blazars.  These two objects have some interesting 
 observation properties. They are QSOs showing properties similar to those
 of a BL Lacertae object: Large amplitude variation and high and variable
 polarization.  4C29.45 is classified as a BL Lac object (see
 Fan et al. 1993 and references therein), but its line spectrum
 looks like that of a normal QSOs when the continuum is faint 
 ( B $\sim$ 18) (Wills et al. 1983).  3C279 and 3C 345 occupy the same
 place as radio selected BL Lacertae objects (RBLs) in the 
 polarization-Doppler factor diagram and are more like  RBLs than QSOs
 (Fan 1998).
 
 Before $\gamma$-ray observations were available, Impey \& Neugebauer 
 (1988) found that the infrared emission  (1-100 $\mu m$) dominate the 
 bolometric luminosities of blazars. The infrared emissions are also an 
 important component for the luminosity even the $\gamma$-ray emissions 
 are included ( von Montigny 1995). The high energetic $\gamma$-ray
 emission mechanism is still an open problem (see Fan et al. 1998a, and
 references therein). Recently, Xie et al. (1997)
 found that the high energetic $\gamma$-rays are correlated with the
 near-IR emissions suggesting that the soft photons scattered to the
 $\gamma$-rays region are from the dust. So, study of the infrared will 
 throw some lights on the understanding of the emission mechanisms in
 AGNs.

 Blazars have been observed in infrared region for more than 20 years. 
 But there  are no available long-term infrared variations in the literature 
 for all these objects.  In this paper, we mainly present the long-term 
 infrared (J,H, and K bands ) light curves for 3C279 and 4C29.45 and 
 discuss the variation properties because there are more continuous
 observations in their J, H, and K bands.  The paper has been arranged as
follows: 
 In section 2, we present the literature for the data and the light curves;
 in section 3, we discuss them and give a brief conclusion.

\section{Near-infrared Light Curves}

\subsection{Data}

 Infrared observations are available since the 1960s. Here we compiled 
 the data from 22 publications listed in table 1, which gives the 
 observers in Col. 1. and the telescope(s) used in Col.2. 

\begin{table}
\caption[]{Literature and telescopes for the data}
\begin{tabular}{|c|c|}
\hline\noalign{\smallskip}
Observer(s) &  Telescope(s)\\ 
\hline\noalign{\smallskip}
  Brown et al (1989b)        &  UKIRT 3.8m \\ \hline
  Garcia-Lario et a l(1989) & TCS 1.5m  \\ \hline
  Gear et al (1985,1986a)    &  UKIRT 3.8m  \\ \hline
  Glassgold et al(1983)     & UKIRT 3.8m \\
                            & Palomer Mt. 5m \\ \hline
  Holmes et al (1984)        &  UKIRT 3.8m  \\ \hline
  Impey et al (1984)        &  UKIRT 3.8m \\ \hline
  Kidger \& Casares (1989)  & TCS 1.5m  \\ \hline
  Kidger et al(1992)       &  TCS 1.5m  \\ \hline
  Landau et al (1986)      &  UKIRT 3.8m; Hale 5m \&\\
                           &   Mount Lemnon 1.5m \\ \hline
  Lepine et al(1985)       &  ESO 3.6m \\ \hline
  Litchfield et al (1994)  &  ESO 2.2m  \\ \hline
  Maraschi et al(1994)     &  Sutherland 1.9m  \\ \hline
  Mead et al (1990)        &  UKIRT 3.8m\\ \hline
  Neugebauer et al(1979)   & Hale 5.0m  \\ \hline
  O'Dell et al (1978)      &  UM/UCSD 1.5m \\ \hline
  Rieke et al (1977)       &  UOA 90inch \& 61 inch \\ \hline
  Roelling et al (1986)    &  UKIRT 3.8m \\ \hline
  Sitko et al (1982)       & UM/UCSD 1.5m  \\ \hline
  Sitko \& Sitko (1991)    &  KPNO 1.3m \& 1.5m \\ \hline
  Smith et al (1987)       &  KPNO 2.1m \\ \hline
  Takalo et al(1992)       & TCS 1.5m\\ \hline
\end{tabular}
\end{table}

\subsection{Light Curves}

 The flux density in the literature has been convected back to 
 magnitude using the original conversion. No dereddening correction 
 was done for the two high Galactic latitude ( $b^{II} = 57^{\circ}.1$ for 
 3C279, $b^{II} = 80^{\circ}$  for 4C29.45) objects. The measurements were
 not always made in three bands, generally, there are more measurements in H
 and K than in J. 
 The light curves are shown in Fig. 1a-c, and Fig. 2a-c for 3C279
 and 4C29.45, respectively.  In the literature, some 
 measurements from 3C279 have relatively large uncertainties( one point
from Neugebauer
 et al. (1979), one from Lepine et al. (1985), and one from Impey et al. 
 (1984) for instance). Fig 1(a)-(c) show that there is good overall
 agreement in the flux variation. The largest amplitude variations are
 $\Delta J$ = 4.57, $\Delta H$ = 4.26, and $\Delta K$ = 4.45 for 3C279
 and  $\Delta J$ = 3.47, $\Delta H$ = 3.82, and $\Delta K$ = 3.97 for 
 4C29.45. Now, we will analyse the correlations between
color index and  color index and magnitude as well.

\subsection{Correlation}

 Because of the uncertainties in the considered data, straight-line
 fitting  with the uncertainties in both coordinates considered is 
 used to deal with the data.
  $$y(x) = a + bx$$ 
 In principle, $a$ and $b$ can be determined  by 
 minimizing the $\chi^{2}$ merit function, i.e. equation (15.3.2)
 in the book by Press et al. (1992)
$$\chi^{2}(a,b)= {\sum\limits_{i=1}^{n} {(y_{i}-a-bx_{i})^{2}w_{i} }} $$
 where $\sigma_{y_{i}}$ and $\sigma_{x_{i}}$ are the $x$ and $y$
 uncertainties for the $i$th point and
 ${\frac{1}{w_{i}}}=\sigma^{2}_{y_{i}}+b^{2}\sigma^{2}_{x_{i}}$.
 Unfortunately, the occurrence of $b$ in the denominator of above  
 $\chi^{2}$ merit function resulting in equation for 
 ${\frac{\partial \chi^{2}}{\partial b}}=0$ nonlinear makes the task 
 of fitting very complex although we can get a formula for $a$ from
  ${\frac{\partial \chi^{2}}{\partial a}}=0$, 
  $$a={\frac{\sum\limits_{i}w_{i}(y_{i}-bx_{i})}{\sum\limits_{i}w_{i}}}$$
 By minimizing the $\chi^{2}$ merit function with respect to $b$ 
 and using the equation for $a$ at each stage to ensure that the minimum 
 with respect to $b$ is also minimized with respect to $a$, we can get $a$ and $b$,
 but finding the  uncertainties, $\sigma_{a}$ and $\sigma_{b}$, in $a$ and $b$ 
 is  more complicated (see Press et al. 1992). So, in the present paper, we 
 only estimated the regression parameters $a$ and $b$ without
 considering the corresponding  uncertainties in $a$ and $b$.

\subsubsection{Color-color relation}

 The data discussed here and in the following section are given  simultaneously in the original literature 
 except for a few several-day-averaged points of 3C279. 
 From the simultaneous observations, we can get that the averaged color
indices are:
 (J-H) = 0.93$\pm$0.12, (J-K) = 1.89$\pm$0.14, and (H-K) = 0.95$\pm$0.09
 for 3C279; 
  (J-H) = 0.84$\pm$0.09, (J-K) = 1.74$\pm$0.13, and (H-K) = 0.91$\pm$0.09
 for 4C29.45. When the straight-line fitting is used to the color indices,  
  (J-K) is found associated with both (J-H) and (H-K) while there is
 almost no correlation between (J-H) and (H-K) as found in RBLs ( Fan \&
Lin
 1998).
 (J-K) = 1.6(J-H) + 0.41 with the correlation coefficient $r = 0.876$ and the
 probability of the relation happening by chance $p $ is almost 0,
 (J-K) = 2.14(H-K) - 0.17 with $r = 0.72$ and $p$ is almost 0
 for 3C279; 
  (J-K) = 1.52(H-K) + 0.45 with $r = 0.62$ and $p = 4.2 \times 10^{-4}$,
  (J-K) = 1.4(H-K) + 0.48 with $r = 0.705$ and $p = 3. \times 10^{-5}$
 for 4C29.45.
 They are shown in Fig. 1d-f and Fig. 2d-f.

\subsubsection{Color-magnitude relation}

 In AGNs, it is found common for the continuous spectrum to change
 with the brightness of the source, generally, the spectrum flattens
 when the source brightens but an opposite behaviour is found in
 some RBLs (see Fan \& Lin 1998, and references therein). But for the 
 spectral index versus flux density correlation, there is a 
 statistic bias (e.g. Massaro \& Trevese 1996). To avoid this bias,
  we discuss the relation between the magnitude in one band and the 
 color index obtained in other two bands. Used the straight-line-fittin,
 following relations are found for 3C279:
 (H-K) = 0.04J + 0.44 with $r = 0.53$ and $p = 5.0 \times 10^{-7}$,
 (J-K) = 0.08H + 0.92 with  $r = 0.72$ and $p$ is near 0.0,
 (J-H) = 0.06K + 0.26 with  $r = 0.72$ and $p = 2.0 \times 10^{-10}$.
 No similar result is found for 4C29.45 (see Fig. 1g-i, Fig. 2g-i).

\section{Discussion}

 Blazars are variable at all wavelengths, variation in optical band
 show some periodicities for those objects with  some hundred years optical
 observations  (see Fan et al. 1998b ).
 Infrared observations have been done more than 20 years, but no 
 long-term variation is available for AGNs except for the works of Neugebeuer
 et al. (1979) and Litchfield et al. (1994), in which they presented infrared observations of  about 8-year for some selected objects. Recently, we found that
  variations in the optical and infrared
 closely associated for PKS 0735+178 (Lin \& Fan 1998)
 and OJ287 (Fan et al. 1998c) indicating these two bands coming
 from the same mechanism. But, it is reasonable that other nonvariable or 
 slowly varying near-IR component, such as the stars in the parent galaxy,
 is present in the spectrum of AGNs.  In this sense, when the source is
 bright, the spectrum is observed to steepen when the source dims, as
 expected from a synchrotron component which experiencing radiative energy
 losses, but when the source dims further, because of the presence of
 the underlying near-IR emission, the spectrum will flattens with the
 source getting faint. Because the underlying near-IR emission affects J
more 
 serious than other two bands, we would expect that there is a clear 
 tendency of spectral flattening with J when J is fainter than a certain 
 magnitude.

 3C279 is a well known member of OVVs, a large optical variation 
 of $\Delta B\geq6.70$ mag (Eachus \& Liller 1975) and a highly
 optical polarization of $P_{opt}=43.3\pm1.3\%$ ( see Scarpa \& 
 Falomo 1997) are reported. It shows a violently optical brightness increase 
 of 2.0 mag during an interval of 24 hours (Webb et al. 1990). The largest 
 amplitude optical variation is greater than the largest amplitude infrared 
 one. The straight-line fitting gives a very significant linear 
 correlation between (H-K) and J, but Fig.1g indicates that  (H-K)
decreases with J when J is fainter
 than 14 mag indicating the spectrum flattens when the source dims, but his 
 tendency does not show up obviously in Fig. 1h or Fig. 1i. The difference
 between the fitting and the plot is from the faint J points with large
 error bars, which play a less important role on the fitting.  If we 
 remove the two  points with large error bar at the bottom right corner, 
 the tendency is not very clear. Nevertheless, it is worth of  noticing 
 and  being discussed with more data. If this tendency is real, it is 
 perhaps from the affection of the underlying galaxy as discussed above, 
 and the near-IR spectrum of 3C279  consists of, at least, two components.
 The underlying galaxy affects J more serious than H or K, so the tendency
 appeared in Fig. 1g does not show up  clearly in Fig. 1h or 1i.

 4C29.45 also shows large amplitude variation ($\Delta m=5.0$, 
 Branly et al. 1996), high and variable polarizations 
 ($P_{IR}=28.06\% $, $P_{opt}$=28\%, Holmes et al. 1984; Mead et al. 
 1990). The largest amplitude variations
 in the infrared are smaller than, but comparable with, that in the
 optical  band. The infrared light curves show two 
 one-year-separating-double-peaked outbursts
 with an interval of 3.2 years.  During the simultaneous infrared observations,
  4C29.45 showed no significant spectral changing when the source was relative
 bright ($\alpha = -0.98\pm0.18$
  when the source dimmed  by $\sim$ 0.9 mag from K = 9.87 faints to
 10.76) during April 5-8, 1981.  But the spectral index obtained 
 at the end of April ($\alpha = -1.39\pm0.08$) showed spectral
 steepening when the source was about 2.5 magnitude fainter
 than it was on April 5 ( Glassgold et al. 1983). There are no continuous
 observations between them.  From the compiled
 long-term data, this association is complex and some data (Smith et al. 1987)
 have relatively large uncertainties, $\sigma = 0^{m}.1$, (see Fig. 2g-i). 
  From the data, we found that the largest variation in J is 
 smaller than
 that in K. The reason is that there are fewer points in J than H and K
 in the literature. Besides, a  weak  correlation of (J-K) increasing with  H ($p \sim 0.5 \%$) can be 
 obtained. For this object, the data are sparse, its variability
properties should be discussed with more observations.
 
 For the color indices, there are correlations between (J-K) and (H-K) and  
 (J-H) as well, but there is almost no correlation between (J-H) and 
 (H-K). We think the reasons are perhaps due to the facts that (1) 
 (J-K) has wider distribution than (J-H) and (H-K) so that (J-H) and (H-K)
 concentrate in a narrow region diluting the correlation, and (2)
 the spectrum deviates from the power law (Fan et al. 1998d).

 We have compiled the infrared light curves for 3C279 and 4C29.45, largest
 amplitude variation, color-color relation and color-magnitude relation
 are investigated. The largest variation in the infrared is smaller than
 that in the optical band, (J-K) is strongly associated with (J-H) for the 
 two objects  while color-magnitude relation is only found for 3C279,
 the (H-K)-J plot suggests that the spectrum of 3C279 consists of, at
 least, two components. No similar results are found for 4C29.45.

 \begin{acknowledgements}{I am grateful to the anonymous referee for his/her 
 suggestions, I thank  Dr. R.G. Lin (CfA, Guangzhou Normal University, China) 
 for his help  with the data and Dr. Y. Copin (CRAL, Observatoire de Lyon,
 France) for his help with the figures. This work is supported by the
 National Natural Scientific Foundation of China and the National
Pandeng Project of China.} 
\end{acknowledgements}
\cite{}

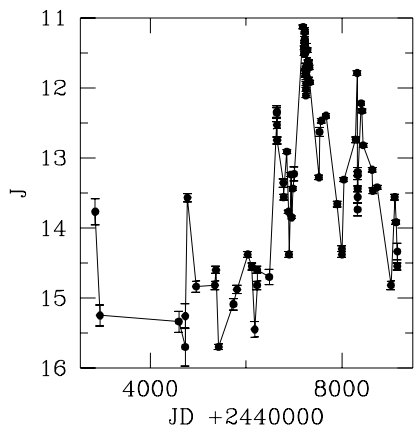
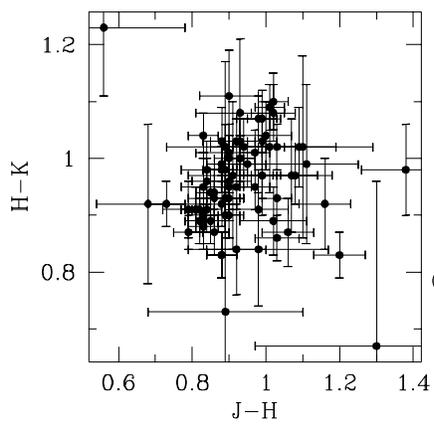
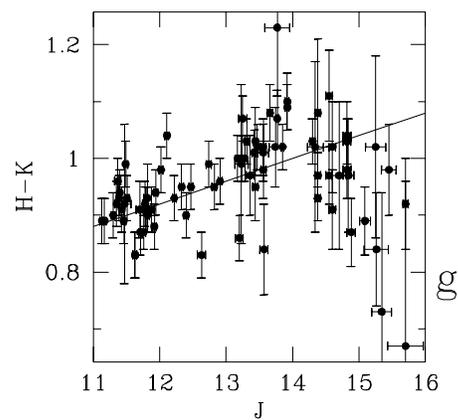
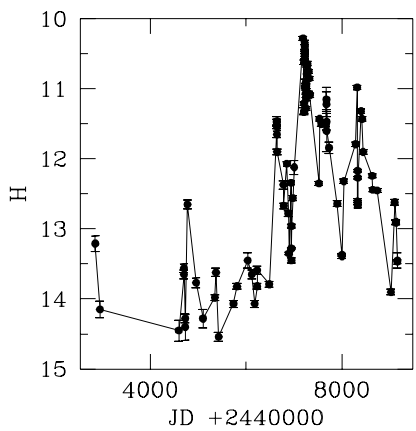
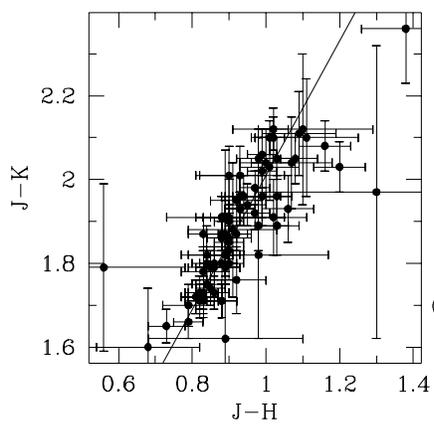
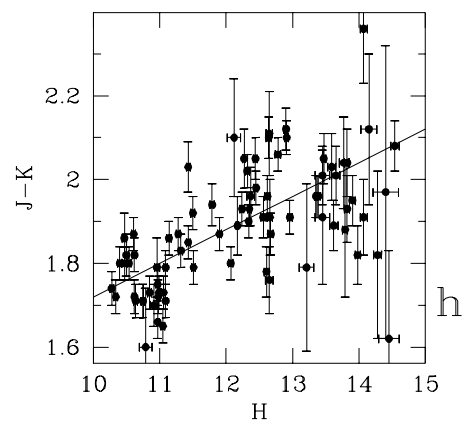
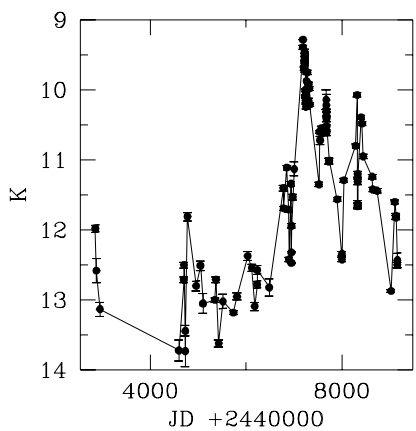
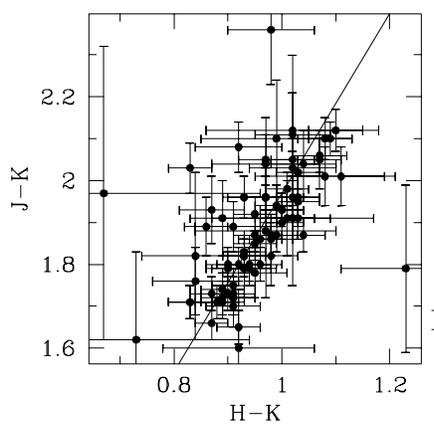
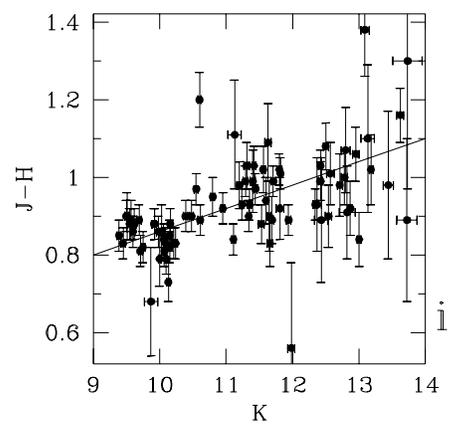
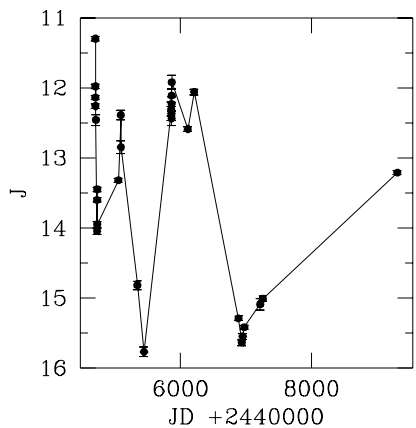
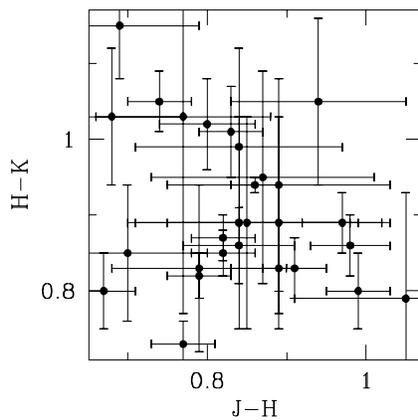
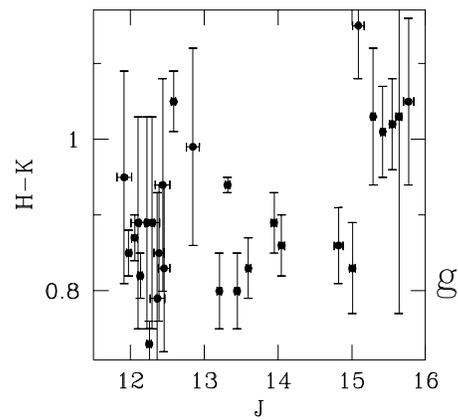
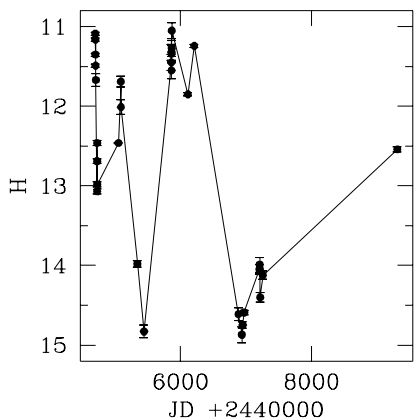
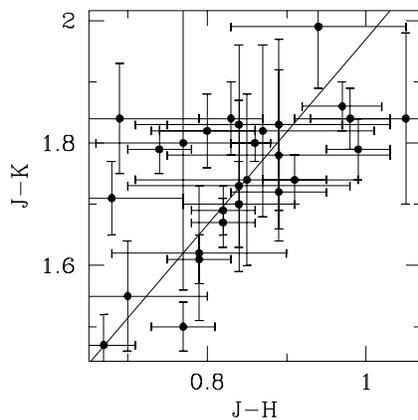
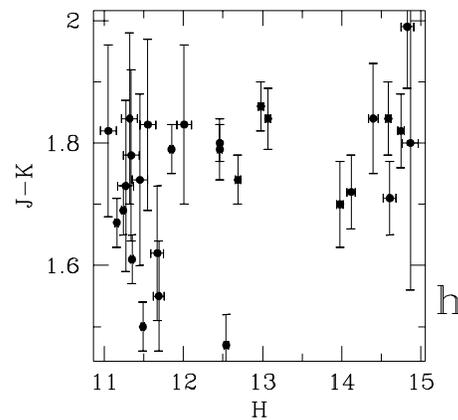
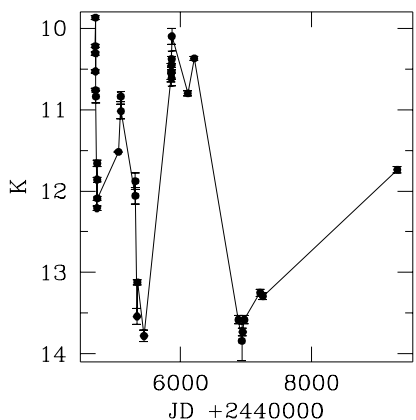
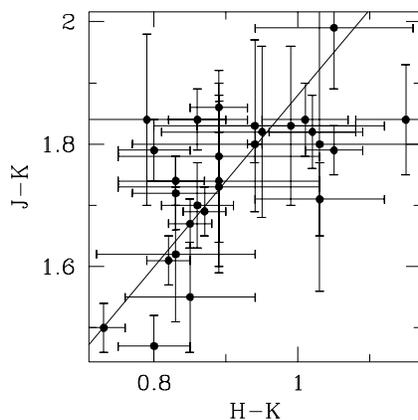
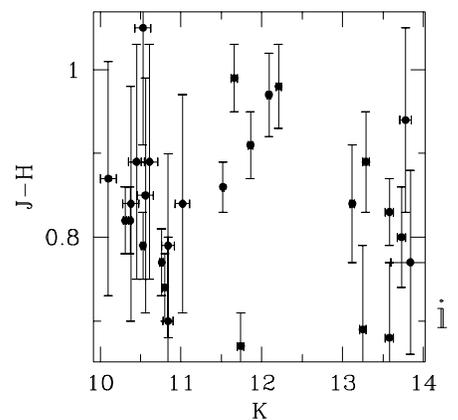
\end{document}